\documentclass[aps,pra,twocolumn,showpacs,floatfix]{revtex4-2}
\usepackage{graphicx}
\usepackage{amsmath}
\usepackage{amssymb}
\usepackage{color}
\usepackage[separate-uncertainty = true,multi-part-units=single]{siunitx}
\newcommand{\Hefour}[0]{$^4\mathrm{He}$}

\begin{document}

\title{Coherent spectroscopy of collective excitations in superfluid helium far from equilibrium}

\author{Gabriel~Voith, Alexander~A.~Milner, Michael~J.~Desrochers, Philip~C.~E.~Stamp and Valery~Milner}

\affiliation{Department of  Physics \& Astronomy, The University of British Columbia, Vancouver, Canada}

\date{\today}

\begin{abstract}
Ultrafast dynamics of collective excitations in superfluids remains largely unexplored beyond the roton region of the Landau excitation spectrum, despite the importance of such dynamics for understanding nonequilibrium processes in these systems. Here, we employ ultrafast coherent control with sequences of femtosecond pulses to perform spectroscopy of multiple quasiparticles in superfluid \Hefour{} far from equilibrium. By measuring the time-resolved optical birefringence, we track the nonequilibrium dynamics of quasiparticle pairs associated with rotons, maxons and the Pitaevskii plateau region. The spectral lineshape of the roton peak is explained by an \textit{ab initio} theoretical analysis of the roton-roton interaction. We also reveal strong energy shifts and short lifetimes of both maxon and Pitaevskii-plateau pairs, as well as an influence of the quasiparticle effective mass on the phase of their coherent response to laser pulses. These results demonstrate the ability to extract previously inaccessible information about collective excitations in a strongly interacting quantum fluid by probing its nonequilibrium dynamics on picosecond and sub-picosecond timescales.
\end{abstract}
\maketitle

Superfluid \Hefour{} (hereafter referred to as ``helium'' for brevity) is one of the paradigm systems for studying collective excitations in strongly interacting many-body quantum systems. Starting with the work of Landau \cite{Landau1941a, Landau1941b}, it is well known that the low-energy excitations in superfluid \Hefour{} consist of a single quasiparticle branch (which includes phonon, maxon, roton, and Pitaevskii plateau sections), coherent quasiparticle pairs, and incoherent multi-particle excitations. Many aspects of superfluidity can be described and explained by the dispersion properties of these quasiparticles and their interactions \cite{Leggett2006, Griffin1993, Nozieres1994}. Until recently, collective excitations in superfluid helium have been studied predominantly with experimental tools sensitive to the equilibrium steady-state dynamics of the superfluid -- neutron scattering \cite{Beauvois2018, Godfrin2021} and spontaneous Raman scattering \cite{Greytak1969, Greytak1970, Ohbayashi1998} (for a recent review, see Ref. \citenum{Glyde2017}). Neither of these approaches can reveal the fast nonequilibrium dynamics of the system: neutron and spontaneous Raman scattering average over timescales far longer than those governing quasiparticle interactions, and can only probe the system in thermal equilibrium.

Recently, we introduced a time-resolved optical method for exciting and tracking roton pairs in superfluid helium on femtosecond and picosecond timescales \cite{Milner2023a, Milner2023b}, exposing rich nonequilibrium dynamics with both the instantaneous frequency and linewidth of the two-roton state evolving rapidly during equilibration with the superfluid bath. Even though the oscillatory birefringence response of the superfluid to an intense laser pulse has recently been treated theoretically in terms of quantum squeezing of quasiparticle pairs \cite{Melnikovsky2026}, the observed asymmetric spectral lineshape of the two-roton peak remained largely unexplained.

While rotons have received the most attention both theoretically and experimentally, the full quasiparticle dispersion curve of superfluid \Hefour{} also includes a local energy maximum associated with maxons, as well as a broad, asymptotically flat region at higher momenta known as Pitaevskii plateau \cite{Pitaevskii1959, Lifshitz1980}. The properties of maxons and their interactions with other quasiparticles are not as well understood as those of rotons. Neutron scattering experiments have established the equilibrium value of the maxon energy gap \cite{Beauvois2018, Godfrin2021}, but the behavior of maxons far from equilibrium, following the rapid injection of a significant amount of energy, remains entirely unexplored. While neutron scattering measurements have established the Pitaevskii plateau energy at twice the roton gap \cite{Glyde1998, Godfrin2021}, several aspects of its interpretation remain debated \cite{Beauvois2018}.

\begin{figure*}[!t]
  \includegraphics[width=.85\textwidth]{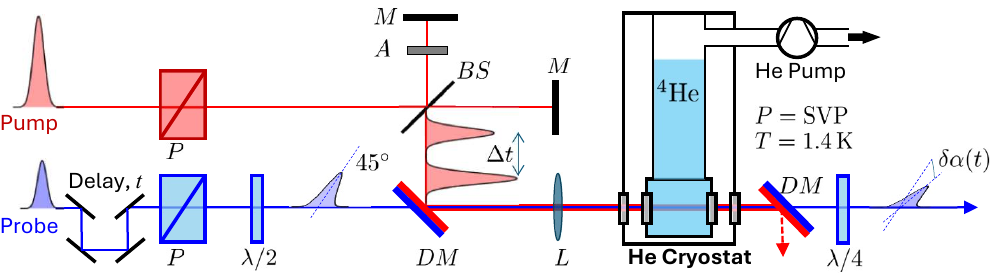}\\
  \caption{Diagram of the experimental setup. Femtosecond pulses with the central wavelength of 798~nm (upper, red) and 399~nm (lower, blue) serve as the pump and probe, respectively. The pulses are linearly polarized at $45^{\circ}$ with respect to one another, combined in a collinear geometry and focused into the bulk liquid helium, kept at the temperature of \SI{1.4}{K} and saturated vapor pressure (SVP). A Michelson interferometer splits pump pulses into pairs with a variable delay $\Delta t$ and relative amplitude. $P$: polarizer, $M$: mirror, $A$: variable attenuator, $BS$: beam splitter, $DM$: dichroic mirror, $L$:lens, $\lambda/2, \lambda/4$: wave plates.}
  \label{fig-Setup}
\end{figure*}
Here, we experimentally record and theoretically analyze the spectrum of collective excitations in superfluid \Hefour{}. From the roton part of the spectrum, our \textit{ab initio} calculations -- a revision of the earlier theory \cite{Ruvalds1970, Zawadowski1972, Bedell1982, Bedell1984} with the phonon-exchange interaction between rotons added on a rigorous footing -- provides the energies and decay rates of bound roton pairs with angular momenta $\ell=0$ and $\ell=2$. We also demonstrate the spectral response of pairs of maxons and Pitaevskii plateau excitations. While hints of these contributions to the birefringence signal are already visible in the single-pulse excitation spectrum, we show that coherent control with sequences of pump pulses greatly facilitates the extraction of their properties, enabling more accurate determination of their energies, linewidths, and phases.

We find that the energy of a maxon pair is lower than twice the single maxon gap by about \SI{40}{GHz} -- significantly larger than the corresponding energy shift for roton pairs. The Pitaevskii plateau contribution also appears at an energy below its expected value. Time-resolved measurements show that maxon pairs decay on a few-picosecond timescale -- an order of magnitude shorter than the lifetime of roton pairs. Finally, our analysis of the birefringence oscillations at two-roton and two-maxon frequencies reveals that the roton and maxon responses are out of phase with one another.

Our experimental setup, shown in Figure~\ref{fig-Setup}, is similar to that described in our previous work \cite{Milner2023a}. Briefly, linearly polarized infrared femtosecond pulses ($\approx$\SI{70}{fs} pulse length, 1~kHz repetition rate, 798~nm central wavelength, intensity $\approx 10^{12}$~W/cm$^2$) are focused in the bulk liquid \Hefour{}, condensed in a custom-built optical cryostat (Lake Shore Cryotronics). By pumping the helium gas from the cryostat, the temperature of the liquid is stabilized at \SI{1.4}{K} at the saturated vapor pressure (SVP). The laser beam passes through a Michelson interferometer, producing pairs of pulses with a variable separation $\Delta t$ and controllable relative amplitude. The pump-induced time-dependent birefringence of the superfluid is detected by measuring the change in the polarization angle $\delta\alpha$ of a time-delayed probe pulse (100~fs pulse length, 399~nm central wavelength), derived from the same laser system and frequency-doubled for easy separation from the pump light. The birefringence signal is recorded as a function of the pump-probe delay $t$ for a given pulse pair separation $\Delta t$.  Since $\delta\alpha$ is small, on the order of \SI{1}{mrad}, sensitive detection is required, which in our case is implemented by modulating the polarization of the pump pulses with a Pockels cell, followed by lock-in amplification (for details, see Ref.~\cite{Milner2023a}).

\begin{figure}[b]
  \includegraphics[width=1\columnwidth]{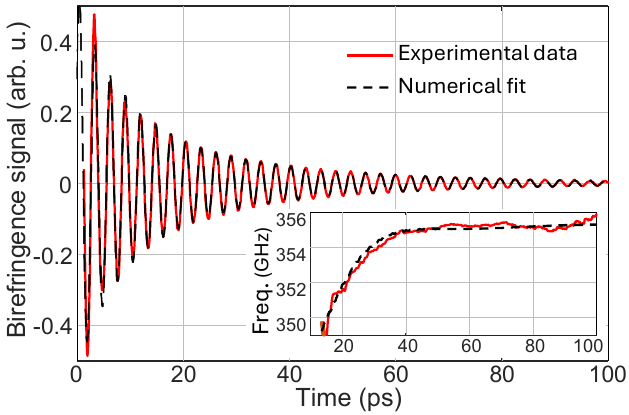}\\
  \caption{Laser-induced optical birefringence of superfluid helium at $T=\SI{1.4}{K}$ and saturated vapor pressure (solid red line). Dashed black line is a fit to the numerical model, based on our \textit{ab initio} theory of the two-roton response. Inset: the dependence of the instantaneous two-roton frequency on time, extracted from the raw data and the numerical model.}
  \label{fig-SingleKickTime}
\end{figure}

Figure~\ref{fig-SingleKickTime} shows the time-dependent optical birefringence of superfluid helium induced by a single femtosecond pump pulse. The oscillatory signal is dominated by the two-roton response around \SI{355}{GHz}, consistent with our earlier observations \cite{Milner2023a}. As reported previously, the instantaneous frequency of this signal is not constant but changes on a picosecond timescale, as depicted in the inset. Our recent observation that the signal amplitude scales linearly with the pump pulse energy \cite{Voith2026b}, with the initial roton frequency remaining unchanged, suggests that the frequency evolution is not driven by laser-induced temperature changes.

\begin{figure}[t]
  \includegraphics[width=.85\columnwidth]{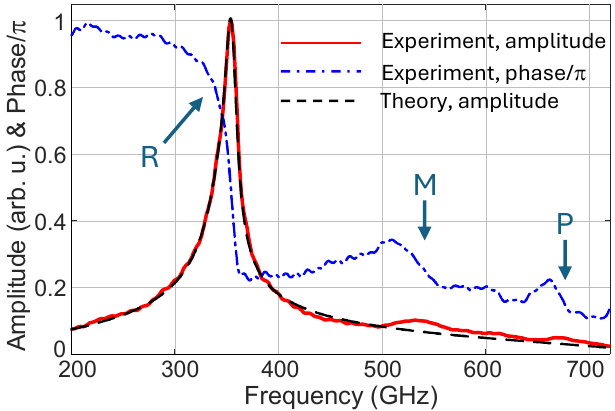}\\
  \caption{Spectral amplitude (solid red) and phase (dot-dashed blue) of the time-dependent signal shown in Fig.~\ref{fig-SingleKickTime}. The three amplitude peaks, accompanied by the corresponding phase steps, are labeled with `R' (two-roton contribution), `M', and `P' (see text for the assignment of the latter two). Dashed black line shows a fit to the theoretical birefringence spectrum, based on the calculated four-point density correlation function $G_{4}(\omega )$ (see Supplemental Material for details).}
  \label{fig-SingleKickSpectrum}
\end{figure}

Instead, we explain the observed roton dynamics by calculating the spectrum of the four-point density correlation function $G_{4}(\omega )$ (see Supplemental Material), which is known to describe Raman scattering (and here, the time-dependent birefringence) in superfluid helium \cite{Halley1969, Stephen1969, Stephen1976, Halley1989}. Our theoretical calculations of $G_{4}(\omega )$, which currently include only the roton response, are in good agreement with the spectral amplitude of the experimental two-roton peak, as shown in Figure~\ref{fig-SingleKickSpectrum}. The fit provides binding energies of $\ell=0$ and $\ell=2$ roton pairs: $E^{r}_{0}/h=\SI{15.0(5)}{GHz}$ and $E^{r}_{2}/h=\SI{6.0(2)}{GHz}$, respectively (where $h$ is the Planck's constant), as well as their decay rates: $\gamma^{r}_{0}=\SI{11.0(5)}{GHz}$ and $\gamma^{r}_{2}=\SI{6.0(5)}{GHz}$. The properties of the $\ell=2$ two-roton state, calculated here from first principles, are consistent with the previously known estimates \cite{Murray1975, Bedell1984, Shay2007}. Our fit also indicates that the short-range contact interaction between rotons is repulsive, with a magnitude of $g_{4}=\SI{30.0(5)}{K \AA^{-3}}$.

Owing to the different energies and decay constants of the unbound roton pairs and the two bound $\ell=0,2$ components, their interference gives rise to the observed frequency chirp in the time-domain signal \cite{Milner2023a}. We illustrate this by fitting the raw experimental data in Fig.~\ref{fig-SingleKickTime} with a sum of three decaying oscillations at fixed frequencies of $2\Delta _{0}=360$~GHz and $2\Delta _{0}-E^{r}_{0,2}/h$, where $h \Delta _{0}$ is the free roton energy known from the neutron scattering experiments \cite{Pearce2001}. Good quality of the fit, shown in Fig.~\ref{fig-SingleKickTime}, as well as a good agreement between the corresponding instantaneous frequency dynamics (inset), confirms the validity of the ``three component'' numerical model for the roton response, based on theoretically calculated frequencies and decay rates.

Aside from the dominating two-roton peak in Figure~\ref{fig-SingleKickSpectrum} (`R'), two other weak features are visible at higher frequencies, located at around twice the maxon energy (`M') and twice the energy of the Pitaevskii plateau (`P'). The existence of these weak resonances is corroborated by the spectral phase, which exhibits a characteristic sharp step across each of the corresponding frequencies -- a hallmark of a resonant contribution to the signal. We notice that the `M' and `P' peaks ride on the shoulder of the two-roton peak, which is one to two orders of magnitude stronger, making it difficult to reliably extract their precise frequencies and linewidths. We overcome this limitation by implementing the double-pulse excitation scheme, described below.

\begin{figure}[t]
  \includegraphics[width=.8\columnwidth]{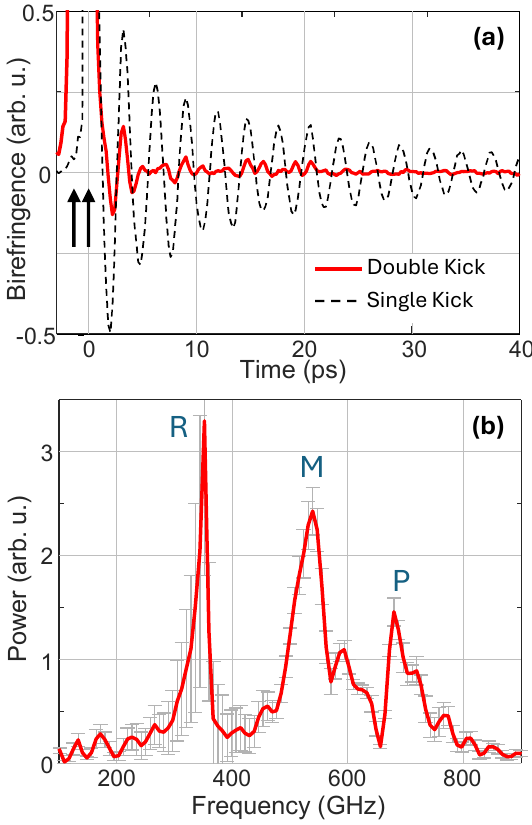}\\
  \caption{(\textbf{a}) Birefringence signal (red solid line) recorded after the excitation by two pump pulses, separated in time by $\Delta t =\SI{1.45}{ps}$ (as indicated with two vertical arrows). For reference, the response to the second pulse alone (analogous to that in Fig.~\ref{fig-SingleKickTime}) is shown by the dashed black curve. The power spectrum of the residual signal is shown in panel (\textbf{b}), with the `R', `M', and `P' peaks labeled as in Fig.~\ref{fig-SingleKickSpectrum}.}
  \label{fig-DoubleKick}
\end{figure}
The key to isolating the weak `M' and `P' peak contributions from the dominant roton signal lies in the well-known technique of coherent control \cite{ShapiroBrumerBook}. It exploits the linearity of the system's response to the laser field: when two excitation pulses are applied in succession, the total response is a superposition of the two individual responses, and their relative phase can be tuned to achieve destructive interference, effectively annihilating the wave packet associated with a particular excitation. Though widely employed for controlling molecular dynamics (e.g., to annihilate a rotational wave packet in molecules \cite{Lee2006}), to the best of our knowledge, selective suppression of one collective excitation for the purpose of isolating and studying another one in a strongly interacting many-body environment has been recently suggested theoretically \cite{Machado2025}, but never implemented experimentally.

Figure~\ref{fig-DoubleKick}(a) shows the time-dependent birefringence recorded after the excitation by two pump pulses separated by $\Delta t = \SI{1.45}{ps}$, chosen such that the two roton responses interfere destructively. The degree of suppression can be inferred by comparing the residual signal (solid red) with the one from the second kick alone (dashed black). As seen in the plot, the suppression of the roton signal is incomplete. This is partly due to the three-component nature of the roton response discussed above, which prohibits full annihilation with a double-pulse sequence. More importantly for the present work, collective excitations oscillating at frequencies distinct from that of rotons are not suppressed by the roton-targeted destructive interference, and their contributions become more prominent in the residual signal.

In the experiment, we iteratively vary both the temporal separation $\Delta t$ and the relative amplitude of the two pump pulses to maximize the relative weight of the weak spectral features. The resulting spectrum is shown in Figure~\ref{fig-DoubleKick}(\textbf{b}). Both the maxon peak and the one corresponding to the Pitaevskii plateau are much more pronounced than in the single-pulse response, allowing us to analyze their properties with higher accuracy.

Similarly to the optical excitation of rotons, conservation laws dictate that other quasiparticles are created with zero total momentum \cite{Halley1969, Stephen1969}. Hence, the `M' resonance frequency of \SI{540(10)}{GHz} may correspond to either a three-roton bound state \cite{Hirashima1989} at $\approx3\Delta _0$, or maxon pairs whose frequency is about \SI{35}{GHz} below $2\Delta _1$, where $\Delta_1=\SI{288}{GHz}$ is the maxon energy at our temperature and pressure, determined by the high-resolution neutron scattering technique \cite{Gibbs1999, Beauvois2018, Godfrin2021}. Being nearly degenerate, the two states must also couple, both adding to the observed peak.

We assign the dominant contribution to the `M' peak to maxon pairs on the following grounds. Our data show that between 1.44 and \SI{1.64}{K}, the width of this peak remains unchanged, whereas the two-roton peak broadens by more than a factor of two over the same range, largely due to the expected increase of the roton linewidth (see Supplemental Material). Since the maxon linewidth is known to depend much more weakly on temperature \cite{Gibbs1999}, a feature composed of rotons should broaden with the two-roton peak, whereas one composed of maxons should not. Moreover, the two-maxon energy downshift is not unexpected. Quasiparticle interactions remove the square-root singularity in the density of states and shift spectral weight into the two-maxon continuum, which extends downward from $2\Delta _1$ owing to the negative curvature of the dispersion \cite{Zawadowski1972}.

The power spectrum of the residual signal also reveals a peak at \SI{680(10)}{GHz}, which is about \SI{40}{GHz} lower than twice the Pitaevskii plateau energy, known from the neutron scattering experiments. There, the plateau is found at $2\Delta _0\approx \SI{360}{GHz}$ \cite{Glyde1998, Godfrin2021}, which would correspond to $4\Delta_0 \approx \SI{720}{GHz}$ in our experiment, owing to the pairwise excitation of quasiparticles. By analogy with the `M' peak above, we attribute the spectral feature at \SI{680(10)}{GHz} mainly to pairs of quasiparticles at the Pitaevskii plateau: the deficit is consistent with the interaction-induced displacement of spectral weight into the pair continuum. This peak is, however, too weak to permit the corresponding temperature analysis.

The power spectrum shown in Figure~\ref{fig-DoubleKick}(b) also allows us to extract the linewidth of the maxon pair resonance. The measured value of approximately \SI{70}{GHz} is about an order of magnitude broader than both the $\ell=0$ and $\ell=2$ two-roton linewidths,  indicating a significantly faster decay of maxon pairs. We also note the asymmetric shape of both the `M' and `P' peaks, which, similarly to the skewness of the roton line shape, may reflect the inhomogeneous nature of those collective modes, e.g. owing to the simultaneous excitation of unbound and bound pairs of quasiparticles with different angular momenta.

\begin{figure}[!t]
  \includegraphics[width=0.9\columnwidth]{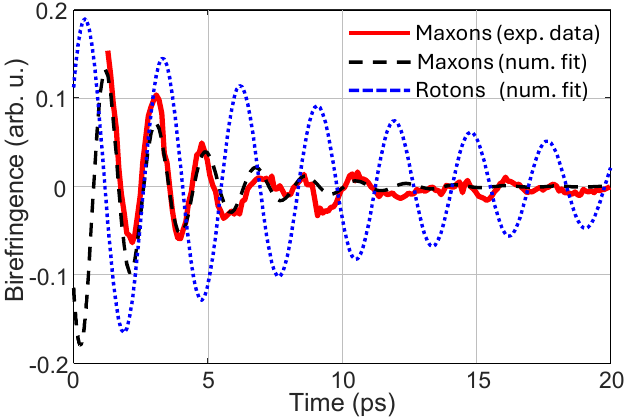}\\
  \caption{Time-dependent contribution of maxon pairs to the measured birefringence (solid red line), obtained by subtracting the three-component numerical model of the roton response (blue dotted line, scaled down by a factor of 2.7) from the total signal. The black dashed curve represents the fit to the maxon-pair contribution at \SI{536}{GHz} (see text for details).}
  \label{fig-Phase}
\end{figure}
Armed with the three-component theoretical description of the roton response, and with the maxon pair frequency extracted from the double-pulse experiment, we proceed with the analysis of the relative phase between the roton and maxon contributions to the laser-induced birefringence. Maxon oscillations are isolated by subtracting the numerical model of the roton response, backed by our theoretical calculations, from the overall birefringence signal (dashed black and solid red lines in Fig.~\ref{fig-SingleKickTime}, respectively). The result is shown in Figure~\ref{fig-Phase} as a solid red curve. This allows us to independently determine the decay constant and phase of the maxon response. The value of the former, $\SI{3(1)}{ps}$, is somewhat shorter than \SI{4.5}{ps} expected from the linewidth of the two-maxon peak of $\approx\SI{70}{GHz}$. This again points at the possibility of an inhomogeneous nature of the laser-induced maxon pairs.

The extracted phase reveals that the roton and maxon responses clearly start out of phase: when described by a single oscillation close to $t=0$, the two-roton phase of $\SI{0.5(2)}{rad}$ and that of the maxon response ($\SI{-2.5(5)}{rad}$) are quite different, as illustrated by the dashed blue and black curves in Figure~\ref{fig-Phase}, respectively. We note that the apparently more straightforward approach of extracting the maxon phase directly from the double-pulse residual signal (Fig.~\ref{fig-DoubleKick}) is in fact unreliable, as at $\Delta t = \SI{1.45}{ps}$, both the roton and maxon responses to the first pulse have not yet decayed, and the second pulse may therefore alter their relative phase.

In summary, the ultrafast dynamics of bound and unbound roton pairs, induced by a sudden injection of energy with a femtosecond laser pulse, has been explained by an \textit{ab initio} theoretical model, which incorporates a phonon-exchange interaction potential between a pair of rotons (see Supplemental Material). Fitting the experimental spectrum of the laser-induced birefringence to the calculated four-point density correlation function $G_{4}(\omega )$ offered the binding energies of roton pairs with an angular momentum of $\ell=0$ (previously unknown) and $\ell=2$ (consistent with the earlier estimates), together with their decay rates.

By applying the method of coherent control -- specifically, the destructive interference of the material response to two successive pump pulses -- we demonstrate a new approach to the spectroscopy of collective excitations in superfluid helium far from equilibrium. Selectively suppressing the dominant roton response revealed weaker spectral features that are otherwise obscured, enabling optical observation of maxon pairs and, more tentatively, pairs of quasiparticles associated with the Pitaevskii plateau.

Our results show that the frequency shift of the maxon pair below twice the single-maxon energy is about an order of magnitude larger than the corresponding shift for roton pairs. Similarly, the maxon pair lifetime is about an order of magnitude shorter. The Pitaevskii plateau peak also appears below the value of $4\Delta_0$ inferred from neutron scattering data. The energy deficits of both peaks are consistent with the interaction-induced displacement of spectral weight into the pair continuum, while the temperature independence of the maxon linewidth between 1.4 and \SI{1.6}{K} supports its assignment predominantly to pairs of maxons rather than to a multi-roton state. The interpretation of asymmetric lineshapes of both peaks calls for further experimental and theoretical investigation.

Finally, the relative phase analysis of the roton and maxon responses provides new insights into the excitation mechanism of collective excitations by ultrashort laser pulses in superfluid helium. Although the accuracy of the extracted maxon phase is limited to $\pm\SI{0.5}{rad}$, it is clearly distinct from the roton phase. This is qualitatively consistent with the theoretical prediction that the phase of each quasiparticle's contribution to the birefringence signal is determined by the sign of its effective mass at the corresponding stationary point of the dispersion curve -- positive for rotons and negative for maxons \cite{Lifshitz1980, Melnikovsky2026}. The exact value of the phase difference is likely an intricate function of the quasiparticle interactions, whose importance is revealed in this work.

The reported results represent a further step towards the much needed understanding of the nonequilibrium many-body dynamics of superfluid helium.

The authors thank L.~Melnikovsky and I.~MacPhail-Bartley for valuable discussions. This research was supported by the Natural Sciences and Engineering Research Council of Canada (NSERC).

\section*{Supplemental Materials}
\subsection*{Microscopic Theory of Superfluid \Hefour{}}

The microscopic theory of Helium-4 superfluid can be written in terms of quasiparticles in a way exactly analogous to the treatment of Fermi liquids in the Landau Fermi-liquid theory  \cite{Abrikosov1975, Lifshitz1980, Ruvalds1970, Bedell1984}. It begins with the Bethe-Salpeter equations for renormalized quasiparticles which can be written in short-hand form as

\begin{equation}
\Gamma=\mathcal{I}+\mathcal{I}\mathcal{G}\mathcal{G}\Gamma, \label{eq:Bethe-Salpeter1}
\end{equation}
where we have suppressed all integrations and indices and where $\mathcal{I}$ is the irreducible four-point vertex in the particle-particle channel, $\mathcal{G}$ is the exact quasi-particle propagator, and $\Gamma$  is the exact four-point vertex, describing interactions between quasi-particles.

In the long-wavelength limit, it is possible to describe the dynamics of density fluctuations of superfluid Helium-4 by considering the long-wavelength action \cite{Popov1965, Popov1972},
\begin{equation}
S=\int dt\int d^{D}r\left[\kappa_{0}\rho\left(\mathbf{r},t\right)\dot{\Phi}\left(\mathbf{r},t\right)-\mathcal{H}\left(\mathbf{r},t\right)\right],
\label{eq:Hydrodynamic Action}
\end{equation}
where $\kappa_{0}=\frac{h}{m_{4}}$ is the quantum of circulation, $m_{4}$ is the mass of the Helium-4 atom, $\Phi\left(\mathbf{r},t\right)$ is a field representing the phase of the superfluid (which gives rise to the velocity vector field through $\mathbf{v}\left(\mathbf{r},t\right)=\frac{h}{m_{4}}\nabla\Phi\left(\mathbf{r},t\right)$)
and $\rho\left(\mathbf{r},t\right)$ is the superfluid density. $\mathcal{H}\left(\mathbf{r},t\right)$ is the Hamiltonian density of the long-wavelength theory and is defined as \cite{Desrochers2026},
\begin{equation}
\mathcal{H}\left(\mathbf{r},t\right)=\frac{1}{2}\left(\frac{\kappa_{0}}{2\pi}\right)^{2}\rho\left(\mathbf{r},t\right)\left|\nabla\Phi\left(\mathbf{r},t\right)^{2}\right|-\epsilon\left(\mathbf{r},t\right),
\label{eq:Hydrodynamic Hamiltonian Density}
\end{equation}
\begin{align}
\epsilon\left(\eta,\nabla\eta\right) & =\frac{\kappa^{2}_{0}}{8\rho}\left|\nabla\eta\left(\mathbf{r},t\right)\right|^{2}+\frac{1}{2\chi\rho^{2}_{\text{s}}}\eta^{2}\left(\mathbf{r},t\right)\nonumber\\
 & +\mathcal{O}\left(\eta^{4},\left|\nabla\eta\left(\mathbf{r},t\right)\right|^{4}\right),
\label{eq:density fluctuations}
\end{align}
where quantum fluctuations of $\Phi\left(\mathbf{r},t\right)$
and $\rho\left(\mathbf{r},t\right)$ about the background superfluid phase $\Phi_{0}\left(\mathbf{r},t\right)$
and the background superfluid density $\rho_{\text{s}}\left(\mathbf{r},t\right)$ are written as
\begin{align}
\hat{\Phi}\left(\mathbf{r},t\right) & =\phi_{0}\left(\mathbf{r}\right)+\hat{\phi}\left(\mathbf{r},t\right),
\label{eq:Phi}\\
\hat{\rho}\left(\mathbf{r},t\right) & =\rho_{\text{s}}\left(\mathbf{r}\right)+\hat{\eta}\left(\mathbf{r},t\right).
\label{eq:Rho}
\end{align}

\begin{figure*}
\centering{}\includegraphics[width=0.6\paperwidth]{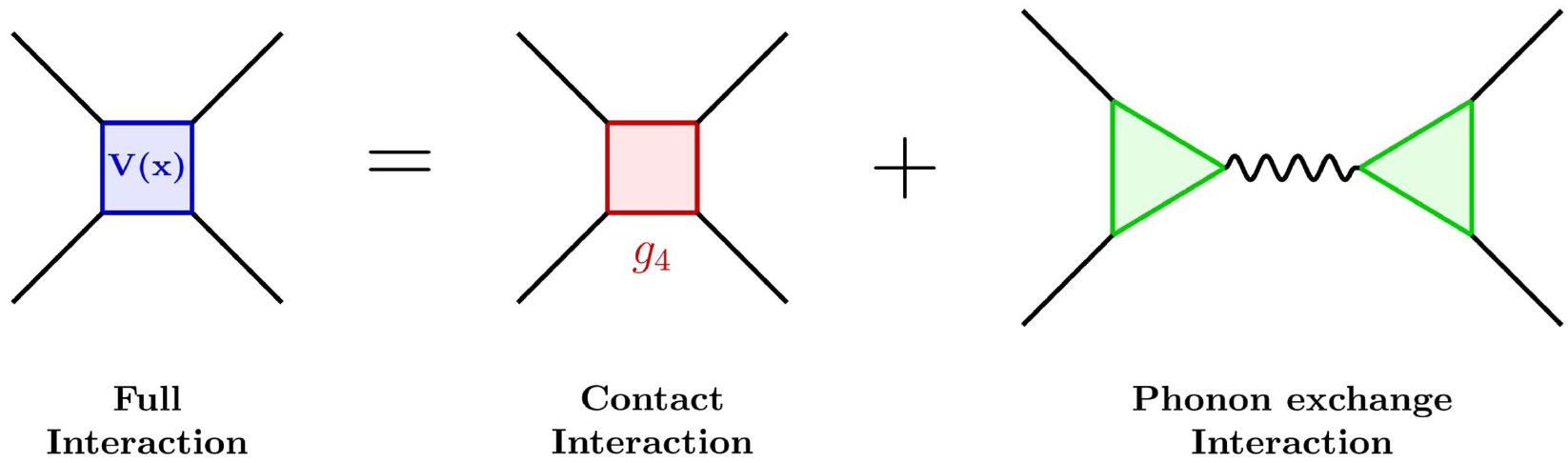}
\caption{\label{fig:Interaction} The full interaction between high-momentum
density fluctuations is represented here in terms of their respective
Feynman diagrams. The contact interaction appears at first order in
perturbation theory while the phonon-exchange appears at second order.}
\end{figure*}

This framework captures the long wavelength quasiparticle degrees of freedom of the superfluid (phonons). The theory is valid for wavelengths much larger than
the healing length $\xi_{0}\sim1\text{\AA\ }$ and for energy scales
much less than $\Lambda_{\text{UV}}=m_{4}c^{2}_{1}\simeq\SI{146}{GHz}$,
where $c_{1}$ is the speed of sound in the superfluid.

For a description at shorter wavelengths, we write everything in terms of a Bose field $\psi\left(\textbf{r},t\right)$ whose excitations are fully renormalized quasiparticles with dispersion $\epsilon_{\textbf{p}}$. These quasiparticles are described by one-particle propagators $\mathcal{G}_{\textbf{p}}\left(E\right)$, which have the usual form
\begin{equation}
\mathcal{G}_{\textbf{p}}\left(E\right) = \frac{Z_{\textbf{p}}}{E-\epsilon_{\textbf{p}}-\Sigma_{\textbf{p}}\left(E\right)},
\end{equation}
where $E$ is the complex energy, $Z_{\textbf{p}}$ is the quasiparticle wave-function renormalization, and $\Sigma_{\textbf{p}}\left(E\right)$ is the quasiparticle self-energy.

\subsection*{Quasiparticle-Quasiparticle Interactions}

The interactions between quasiparticles will be described in the usual way using the Bethe-Salpeter equation \cite{Abrikosov1975, Lifshitz1980} for the renormalized quasiparticles written as Equation~\ref{eq:Bethe-Salpeter1}. The irreducible vertex part $\mathcal{I}$ can always be split into short-range and long-range parts, where the long-range part is mediated by phonons. The long-range phonon exchange between quasiparticles has long been understood to produce an attractive dipolar interaction \cite{Cohen1957, Feynman1954}, mediated by phonons.

We therefore approximate the irreducible quasiparticle interaction $\mathcal{I}$ by a sum of short-range and long-range interactions,
\begin{equation}
\mathcal{I}_{\mathbf{P}_{2},\mathbf{P}_{2}}\left(\mathbf{q},\omega\right)=g_{4}+V^{\text{ph}}_{\mathbf{P}_{1},\mathbf{P}_{2}}\left(\mathbf{q},\omega\right),
\end{equation}
and write the Fourier transform of $V^{\text{ph}}_{\mathbf{P}_{1},\mathbf{P}_{2}}\left(\mathbf{q},\omega\right)$ in the static limit, where $\omega\rightarrow 0$, as

\begin{widetext}
\begin{equation}
\begin{split}V^{\text{ph}}_{\mathbf{P}_{1},\mathbf{P}_{2}}\left(\mathbf{r}\right)= & -\frac{2}{\hbar c_{0}}\frac{\alpha^{2}}{9}\Lambda^{2}_{\text{p}}\psi^{4}_{\text{r\textbackslash m}}\left[P^{2}_{1}P^{2}_{2}\delta^{3}\left(\mathbf{r}\right)+\frac{\left(P^{2}_{1}+P^{2}_{2}\right)}{4}\nabla^{2}\delta^{3}\left(\mathbf{r}\right)+\frac{1}{16}\nabla^{4}\delta^{3}\left(\mathbf{r}\right)\right]\\
 & +\frac{\alpha^{2}}{9}\psi^{2}_{\text{p}}\frac{\hbar^{4}}{8\pi\hbar c_{0}}\left[\frac{3\left(\mathbf{P}_{1}\cdot\hat{\mathbf{r}}\right)\left(\mathbf{P}_{2}\cdot\hat{\mathbf{r}}\right)-\mathbf{P}_{1}\cdot\mathbf{P}_{2}}{r^{3}}-\frac{4\pi}{3}\left(\mathbf{P}_{1}\cdot\mathbf{P}_{2}\right)\delta^{3}\left(\mathbf{r}\right)\right],
\end{split}
\end{equation}
\end{widetext}
where $\alpha=\frac{1}{2}\left(\frac{\kappa_{0}}{2\pi}\right)^{2}$, and where $\mathbf{P}_{1}$ and $\mathbf{P}_{2}$ denote the incoming momenta of the interacting quasi-particles. The subscripts `m' and `r' indicate whether the quasiparticle fields correspond to a maxon or roton, respectively. The parameters $\psi_{\text{p}}$,$\psi_{\text{r\textbackslash m}}$
are all known functions of the dispersion curve. Hence, this long-range phonon exchange interaction requires no phenomenological low-energy constants. However, the short-range interaction $g_{4}$, which in this approximation is simply a delta function representing the hard-core repulsion, is a free parameter in our calculation.

Away from the static limit, the irreducible interaction $\mathcal{I}$ is a dynamic retarded potential. Taking the static limit reduces the potential to the form given above.

\subsection*{Birefringence Spectrum}

The experimental birefringence power spectrum is proportional to $\left|G_4\left(\omega\right)\right|^{2}$ \cite{Halley1969, Stephen1969, Stephen1976, Halley1989}, where the four-point density correlation function $G_4\left(t\right)$ is defined as
\begin{equation}
G_{4}\left(\mathbf{r},t\right)=\left\langle 0\left|\hat{T}\left\{ \rho\left(\mathbf{r},0\right)\rho\left(\mathbf{r},0\right)\rho\left(\mathbf{r},t\right)\rho\left(\mathbf{r},t\right)\right\} \right|0\right\rangle,
\end{equation}
and can be calculated directly from the four-point vertex $\Gamma$. The equation of motion for $G_{4}$ is thus derived from the Bethe-Salpeter equation \cite{Salpeter1951}. One finds that
\begin{equation}
G^{\ell}_{4}(\omega)=\frac{2F(\omega)}{1-V^{\ell}F(\omega)},
\label{eq:Bethe-Salpeter}
\end{equation}
where $\ell$ denotes the specific partial-wave coefficient in the expansions,
\begin{align}
G_{4}\left(p_{0},\theta,\omega\right) & = \sum_{\text{\ensuremath{\ell}}}\left(2\ell+1\right)G^{\ell}_{\text{4}}\left(p_{0},\omega\right)P_{\ell}\left(\cos\left(\theta\right)\right),\\
V\left(p_{0},\theta\right), & = \sum_{\text{\ensuremath{\ell}}}\left(2\ell+1\right)V^{\ell}\left(p_{0}\right)P_{\ell}\left(\cos\left(\theta\right)\right),
\end{align}
in which $\theta$ is the scattering angle between the two quasiparticles, and $p_{0}\in\left\{ p_{\text{R}},p_{\text{M}}\right\} $ represents the momentum at the roton minimum or maxon maximum, respectively. The function $F\left(\omega\right)$ is the two-particle roton propagator $\mathcal{G}\mathcal{G}$ in the Bethe-Salpeter equation \cite{Griffin1993}. We write $F\left(\omega\right)$ in the form
\begin{equation}
F\left(\omega\right) =F_{\text{0}}\left(\omega\right)+C_{\text{UV}},
\end{equation}
where
\begin{equation}
F_{0}\left(\omega\right)=\frac{i}{(2\pi)^{4}}\int d^{3}k\int^{2\Delta_{\text{M}}}_{0}d\tilde{\omega}G_{\text{R}}(\tilde{\omega},\mathbf{k})G_{\text{R}}(\omega-\tilde{\omega},\mathbf{K}-\mathbf{k}),
\end{equation}
and where $C_{UV}$ represents the contribution to $F\left(\omega\right)$ from high-energy incoherent excitations in $\mathcal{G}$.

The pole structure of $G_{4}(\omega)$ dictates both the bound-state energy levels (from the real part of the pole) and the decay rates $\gamma$ (from the imaginary part of the pole) for each partial wave $\ell$.

\subsection*{Bound-State Energy Levels and their Decay Constants}

We solve for the partial wave components of the four-point correlation function and fit it to the birefringence data shown in Figure~3 of the main text, as follows:
\begin{equation}
G_{\text{fit}}\left(\omega\right)=A\left| G_{4}\left(\omega\right)\right|^{2}+B,
\end{equation}
where $A$ is a proportionality constant accounting for the amplitude scaling between the optical birefringence signal and the theoretical four-point correlation function, and $B$ is an offset parameter which accounts for the instantaneous Kerr effect \cite{Righini1993, Sushkov2004}.

Notably, this comparison with experiment provides the first empirical estimate of the short-range contact interaction strength $g_{4}=\SI{30.0(5)}{K\AA^{-3}}$, while simultaneously accounting for the fundamental phonon-exchange interactions derived from the long-wavelength theory (Figure~\ref{fig:Interaction}).

The poles of both $G_{4}\left(\omega\right)$ and $\Gamma$ allow us to calculate bound-state energies of quasiparticle pairs in both the $\ell=0$ and $\ell=2$ channels. The s-wave bound-state energies require an estimate of the short-range contact interaction $g_4$, which has been given above. This result allows one to estimate the $\ell=0$ bound-state energy level for the first time. The calculated bound-state energies are
\begin{align}E^{R}_{\ell=0} & \approx\SI{15.0(5)}{GHz} & E^{R}_{\ell=2} & \approx\SI{6.0(2)}{GHz}.
\end{align}

\begin{figure}[t]
\centering{}\includegraphics[width=1\columnwidth]{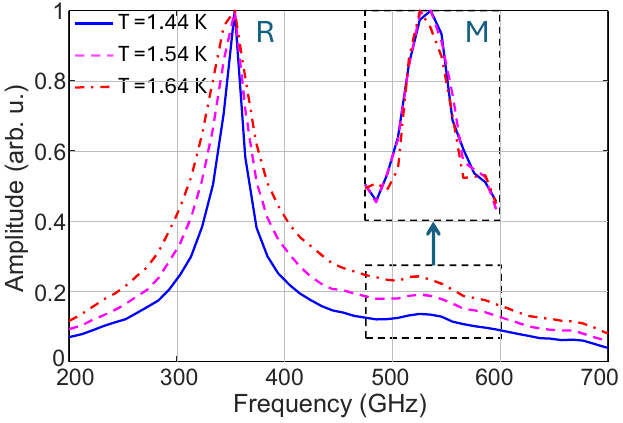}
\caption{\label{fig:TempDependence} Normalized spectral amplitude of the birefringence signal, recorded at \SIlist{1.44;1.54;1.64}{K}. The `M' peak around \SI{530}{GHz} is shown in the inset.}
\end{figure}
Remarkably, without relying on any free fit parameters other than the ultraviolet cutoff $C_{\text{UV}}$ (which parameterizes the high-energy dynamics of the free two-particle propagator), our first-principles calculation of the $\ell=2$ roton-roton bound-state energy agrees well with the estimates derived from the early Raman scattering experiments by Greytak and Yan \cite{Greytak1970}. While previous phenomenological models \cite{Bedell1984} were forced to insert $E^{\text{r}}_{\ell=2}$ by hand to match experimental data, our long-wavelength theory predicts it directly from the phonon-exchange potential \footnote{Even if $C_{\text{UV}}$ is strictly set to zero, the theory yields $E^{\text{r}}_{\ell=2} \approx \SI{6.7}{GHz}$, which remains well within experimental error bounds.}.

A full treatment of the Bethe-Salpeter equation and the pole structure of the four-point correlation function will be detailed in future work \cite{Desrochers2026}.\\

\subsection*{Dependence of the spectral peaks on temperature}
To interpret the nature of the `M' peak, discussed in the main text, we recorded the signal at three different temperatures: \SIlist{1.44;1.54;1.64}{K}. The corresponding spectral amplitudes are shown in Fig.~\ref{fig:TempDependence} with solid blue, dashed magenta, and dash-dotted red curves, respectively. The two-roton peak `R' behaves as expected, shifting to lower frequency and broadening by more than a factor of two, whereas the linewidth of the `M' peak remains unchanged to within roughly 10\% (limited by the signal-to-noise), as illustrated in the inset. The peak is too broad to resolve a frequency shift comparable to the two-roton one. The inset shows the `M' peak after subtraction of the two-roton background, fitted to a linear slope over the frequency window \SIrange{470}{600}{GHz} (using higher-order polynomials does not change the result). This contrast in line broadening supports the assignment of the main contribution to the `M' peak to maxon pairs: a feature composed of rotons would be expected to broaden together with the two-roton resonance, whereas the maxon linewidth is known to depend much more weakly on temperature \cite{Gibbs1999}.



%

\end{document}